\documentclass{article}
\usepackage{spconf,amsmath,graphicx}
\usepackage{marvosym}
\usepackage{spconf,graphicx,tabularray,multirow}
\usepackage{hyperref}
\usepackage{enumitem}
\usepackage{verbatim}
\usepackage{booktabs}
\setlist{nosep, leftmargin=14pt}
\usepackage{multirow}
\usepackage{mwe} 
\usepackage{makecell}

\title{ITCFN: Incomplete Triple-Modal Co-Attention Fusion Network for MILD COGNITIVE IMPAIRMENT CONVERSION PREDICTION}

\name{Xiangyang Hu$^1$, Xiangyu Shen$^1$, Yifei Sun$^1$, Xuhao Shan$^1$, Wenwen Min$^2$, Liyilei Su$^3$,
\\Xiaomao Fan$^3$, Ahmed Elazab$^4$, Ruiquan Ge$^{1,6,*}$, Changmiao Wang$^{5,*}$, Xiaopeng Fan$^6$
\thanks{*Correspondings: gespring@hdu.edu.cn, cmwangalbert@gmail.com.}}
 
\address{$^1$Hangzhou Dianzi University, Hangzhou, China 
$^2$Yunnan University, Kunming, China \\
$^3$Shenzhen Technology University, Shenzhen, China
$^4$Shenzhen University, Shenzhen, China 
\\
$^5$Shenzhen Research Institute of Big Data, China $\,$
$^6$Hangzhou Institute of Advanced Technology, China \\
}

\begin{document}
%
\maketitle
\begin{abstract}
Alzheimer's disease (AD) is a common neurodegenerative disease among the elderly.  Early prediction and timely intervention of its prodromal stage, mild cognitive impairment (MCI), can decrease the risk of advancing to AD. Combining information from various modalities can significantly improve predictive accuracy. However, challenges such as missing data and heterogeneity across modalities complicate multimodal learning methods as adding more modalities can worsen these issues. Current multimodal fusion techniques often fail to adapt to the complexity of medical data, hindering the ability to identify relationships between modalities. To address these challenges, we propose an innovative multimodal approach for predicting MCI conversion, focusing specifically on the issues of missing positron emission tomography (PET) data and integrating diverse medical information. The proposed incomplete triple-modal MCI conversion prediction network is tailored for this purpose. Through the missing modal generation module, we synthesize the missing PET data from the magnetic resonance imaging and extract features using specifically designed encoders. We also develop a channel aggregation module and a triple-modal co-attention fusion module to reduce feature redundancy and achieve effective multimodal data fusion. Furthermore, we design a loss function to handle missing modality issues and align cross-modal features. These components collectively harness multimodal data to boost network performance. Experimental results on the ADNI1 and ADNI2 datasets show that our method significantly surpasses existing unimodal and other multimodal models. Our code is available at \href{https://github.com/justinhxy/ITFC}{https://github.com/justinhxy/ITFC}.

\end{abstract}
\section{Introduction}

Alzheimer's disease (AD) is a progressive neurodegenerative disorder leading to permanent cognitive impairment and eventually death. Cognitive issues are categorized by severity into subjective memory complaints, mild cognitive impairment (MCI), and AD \cite{elazab2024alzheimer}. At the MCI stage, significant cognitive decline is evident, and research indicates that many MCI cases eventually advance to AD. Currently, aside from a few medications that alleviate symptoms, there are no effective treatments to reverse AD or MCI. Thus, early prediction and timely intervention in MCI are crucial to reduce the risk of progression to AD. Progressive MCI (pMCI) describes those MCI patients who likely develop AD within three years, whereas stable MCI (sMCI) patients remain cognitively stable during this time. Accurately predicting the progression of pMCI is vital for early intervention, which can delay the development of AD.  In clinical practice, subjects undergo brain imaging, such as magnetic resonance imaging (MRI) for anatomical assessment and positron emission tomography (PET) for functional evaluation, to assess disease status. Building on this, computer-aided diagnostic systems can be instrumental in distinguishing between pMCI and sMCI at an early stage \cite{abdelaziz2021alzheimer}.

\begin{figure*}[t]
\centerline{\includegraphics[width=0.85\textwidth]{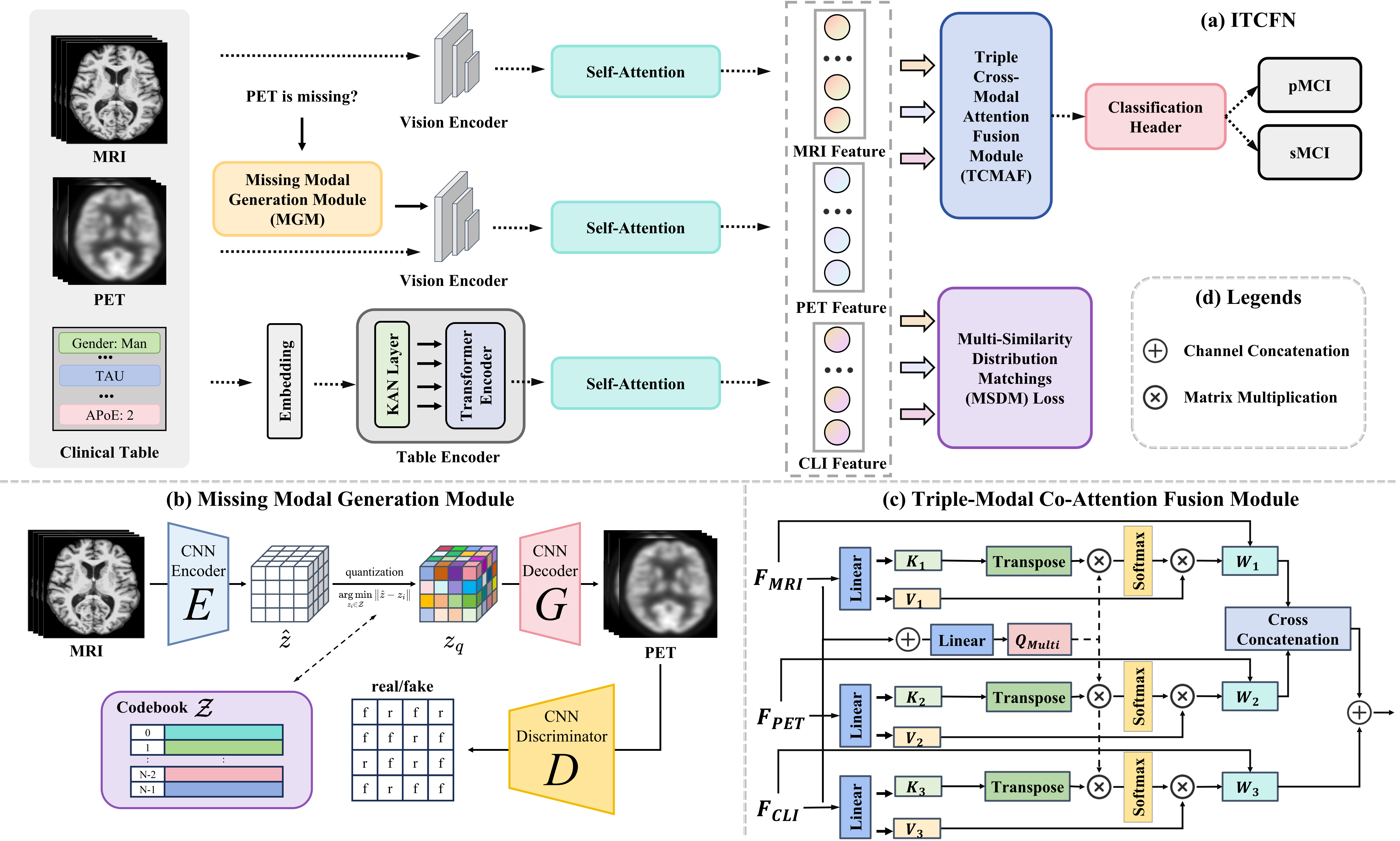}}
\caption{The overall architecture of the ITCFN. (a) Depicting the network flow, including solutions for missing multimodal data, feature extraction, multimodal fusion, and loss function. (b) The MMG module generates missing PET data based on MRI. (c) The TCAF module facilitates multimodal fusion.}
\label{fig:1}
\end{figure*}


Convolutional neural networks (CNNs), like 3D-ResNet \cite{ResNet}, are widely used for diagnosing AD. Despite its success, the progressive stages of AD require careful consideration. Recently, Wang et al. \cite{hope} proposed a HOPE method that focuses on MRI features across different stages and has shown improvements in predicting MCI-to-AD conversion. However, using MRI alone achieves suboptimal performance as fails to capture metabolic changes in the brain, highlighting the limitations of single-modality approaches. Combining MRI and PET modalities is a common strategy since structural and functional neuroimaging provide more holistic information about the brain's status \cite{shoeibi2023diagnosis}. Nevertheless, PET data is often scarce due to costs and radiotracer concerns, resulting in missing data \cite{fang2024gfe, chen2024toward}. To address this limitation, Liu et al. \cite{JSRL} employed a generative adversarial network (GAN) method, JSRL, to generate PET features. However, their fusion method did not fully leverage the complementary information across modalities. There is potential for enhanced performance in multi-modal approaches by exploring diverse features in both imaging and tabular data. While some studies have employed CNNs and multi-layer perceptron as feature encoders. Despite their use, these methods often fail to capture the essential information within each modality, which hampers effective data integration. For instance, Kang et al. \cite{VAPL} presented a Visual-Attribute Prompt Learning transformer-based model (VAPL) to learn MRI and clinical tabular features. Nevertheless, there are still few methods that successfully integrate structural imaging, functional imaging, and clinical tabular data simultaneously.

To address these challenges, we introduce the Incomplete Triple-Modal Co-Attention Fusion Network (ITCFN) for predicting MCI conversion using incomplete multi-modal data. Initially, we propose the Missing Modal Generation (MMG) module to synthesize the missing PET images from the MRI ones. Then, features learned from each modality are extracted using specialized encoders. Finally, these features are combined in a tailored fusion module for MCI conversion prediction. The primary contributions of our study are as follows:

\begin{itemize} 
    \item We develop an MMG module to synthesize the missing PET images, tackling the problem of multimodal data availability.
\end{itemize}
\begin{itemize}
    \item We propose an ITCFN model that effectively learns and integrates multimodal imaging PET and MRI data with clinical tabular data. 
\end{itemize}
\begin{itemize}
    \item We design a new loss function to address data imbalance and align the feature representations across the imaging and non-imaging data.
\end{itemize}
\label{sec:intro}
\vspace{-0.5em}

\section{Method}

\begin{table*}[htbp]
\caption{Characteristics of the datasets used in experiments.}
\centering
\small
\begin{tabular}{|p{1.9cm}<{\centering}|p{1.45cm}<{\centering}|p{1.6cm}<{\centering}|p{1.45cm}<{\centering}|p{1.6cm}<{\centering}|p{1.45cm}<{\centering}|p{1.45cm}<{\centering}|p{1.45cm}<{\centering}|p{1.45cm}<{\centering}|}
\hline
\multirow{2}{*}{Variable} & \multicolumn{4}{c|}{ADNI1 \cite{ADNI}} & \multicolumn{4}{c|}{ADNI2 \cite{ADNI}} \\ \cline{2-9}
& AD & pMCI & sMCI & NC  & AD & pMCI & sMCI & NC   \\ \hline
Number (M/F)  &88/83   &90/61   & 136/72  &103/104   &89/67   &43/38   &156/125   & 132/165     \\ \hline
Age  & 75.35$\pm$7.47 & 74.63$\pm$7.18 & 74.75$\pm$7.63  & 75.92$\pm$5.12&  74.75$\pm$8.09 & 72.60$\pm$7.27 &  71.29$\pm$7.43 & 72.80$\pm$6.01  \\ \hline
Education  & 14.64$\pm$3.19 & 15.66$\pm$ 2.92 & 15.61$\pm$3.11 &  15.91$\pm$2.87&  15.72$\pm$2.75& 16.29$\pm$2.55 & 16.31$\pm$2.61& 16.61$\pm$2.5     \\ \hline
CDR-SB  & 4.32$\pm$1.58  & 1.85$\pm$0.98  & 1.38$\pm$0.75  & 0.03$\pm$0.12& 4.51$\pm$1.67& 2.18$\pm$0.95  &  1.33$\pm$0.82 &  0.04$\pm$0.15      \\ \hline
MMSE  & 23.23$\pm$2.03  & 26.59$\pm$1.7  & 27.33$\pm$1.77  & 29.14 $\pm$0.98 & 23.12$\pm$2.07 & 27.1$\pm$1.82 &28.21$\pm$1.63&  28.99$\pm$1.26    \\ \hline
\end{tabular}
\label{dataset}
\end{table*}

\subsection{Missing Modal Generation Module}
For subjects where the PET modality is missing, we introduce the MMG module to derive PET images from MRI through cross-modal transformation. This module employs a simplified Vector-Quantized Generative Adversarial Network (VQGAN), as illustrated in Fig.~\ref{fig:1}. Initially, a 3D Convolutional Encoder (CNN Encoder) processes the MRI data \(x\) to obtain the vector \(\hat{z}\). The Codebook \(\mathcal{Z}\) is then used to match each coding position of \(\hat{z}\) with its nearest code in \(\mathcal{Z}\), resulting in the variable \(z_{q}\). This intermediate feature \(\hat{z}\) is discretized and encoded as follows:
\begin{equation}
z_{q}=q(\hat{z}):=\underset{z_m \in \mathcal{Z}}{\arg \min }\left\|\hat{z}_{i j k}-z_m\right\|.
\end{equation}
Further, $z_{q}$ is decoded by a 3D Convolutional Decoder (CNN Decoder) to generate the PET image \(\hat{y}\), represented by:
\begin{equation}
\hat{y}=G\left(z_{q}\right)=G(q(E(x))).
\end{equation}
By introducing the vector quantization technology, the uncertainty in the generation process can be effectively reduced, resulting in more stable quality of the generated data.

\textbf{Hybrid Loss Function. }To optimize the training of the MMG module, we devise a hybrid loss function that integrates several components: L1 loss \(\mathcal{L}_{L1}\), quantization loss \(\mathcal{L}_{Qua}\) \cite{tudosiu2022morphology}, perceptual loss \(\mathcal{L}_{Per}\) \cite{johnson2016perceptual}, and adversarial loss \(\mathcal{L}_{Adv}\) \cite{wang2018pix2pixHD}. These components work together to enhance the local realism of the generated PET images. The proposed hybrid loss function is expressed as:
\begin{equation}
\mathcal{L} = \lambda_{L1} \cdot \mathcal{L}_{L1} + \lambda_{Qua} \cdot \mathcal{L}_{Qua} + \lambda_{Per} \cdot \mathcal{L}_{Per} + \lambda_{Adv} \cdot \mathcal{L}_{Adv},
\label{lossvqgan}
\end{equation}
where $\lambda_{L1}$, $\lambda_{Qua}$, $\lambda_{Per}$, and $\lambda_{Adv}$ are the respective weighing parameters for each loss component.

\vspace{-0.5em}
\subsection{Multi-modal Feature Processing}
In this work, we design specific encoders to accommodate the unique characteristics and formats of different data modalities, mapping them into a common feature space for effective fusion. For MRI and PET data, we use the pre-trained 3D ResNet50 as the visual encoder to extract deep features while the clinical data is processed through a custom table encoder to enhance feature representation. The resulting high-dimensional features undergo processing by the multi-head self-attention mechanism with 16 attention heads, enhancing the fusion process's effectiveness, and improving the model's overall performance. 

\vspace{-0.5em}
\subsection{Triple-modal Co-Attention Fusion Module}
The architecture of the Triple-Modal Co-Attention Fusion (TCAF) Module is depicted in Fig.~\ref{fig:1}(c). Modal features are processed through linear layers to obtain the corresponding \(K_i\) and \(V_i\), while \(Q_{multi}\) is derived by concatenating all modalities and mapping them through a linear layer. The attention scores for each modality are calculated as the following equations:
\begin{equation}
    F^{i}_{hidden} = softmax \left( \frac{Q_{multi} K_i}{\sqrt{d_i}} \right) V_i,
\label{eq_att}
\end{equation}
where \(d_i\) is the dimension of the vector \(K_i\).

To fully integrate the intrinsic information of MRI and PET, we employ a cross-concatenation method, followed by concatenation with clinical features to produce the final output using:
\begin{equation}
    F_{global} = \bigoplus^{Cross}(F^{1}_{hidden}, F^{2}_{hidden}) \oplus F^{3}_{\text{hidden}}.
\end{equation}

Eventually, we employ a pre-trained 1D DenseNet-121 as the classifier. The pre-trained parameters enhance model convergence speed and reduce overfitting.

\subsection{Loss Function}
To alleviate the effect of data imbalance, we employ focal loss \(\mathcal{L}_{focal}\), which emphasizes minority and hard-to-classify samples through dynamic weighting and class balancing.
We also utilize the Similarity Distribution Matching (SDM) loss to improve feature alignment across MRI, PET, and clinical data. The SDM losses are computed separately for each modality pair: \(\mathcal{L}_{SDM}^{mt}\) for MRI and Table, \(\mathcal{L}_{SDM}^{pt}\) for PET and Table, and \(\mathcal{L}_{SDM}^{mp}\) for MRI and PET. The combined loss is defined as:

\begin{equation}
        \mathcal{L}_{triple} = \lambda \left( \frac{\mathcal{L}_{SDM}^{mt} + \mathcal{L}_{SDM}^{pt}}{2} \right) + (1 - \lambda) \mathcal{L}_{SDM}^{mp}.
\end{equation}

\textbf{Overall Loss Function. }We utilize a joint loss function to optimize the entire network, combining the task-specific loss with the multimodal alignment loss. This overall loss function is expressed as a weighted sum:
\begin{equation}
   \mathcal{L}_{total} = \mathcal{L}_{focal} + \alpha \mathcal{L}_{triple}.
\end{equation}
\par

\label{sec:format}
\begin{table*}[htbp]
\caption{Comparison performance with previous works.}
\centering
\small
\begin{tabular}{|p{2.7cm}<{\centering}|p{1.2cm}<{\centering}|p{0.8cm}<{\centering}|p{0.8cm}<{\centering}|p{0.8cm}<{\centering}|p{0.8cm}<{\centering}|p{0.8cm}<{\centering}|p{0.8cm}<{\centering}|p{0.8cm}<{\centering}|p{0.8cm}<{\centering}|p{0.8cm}<{\centering}|p{0.8cm}<{\centering}|p{0.8cm}<{\centering}|}
\hline
\multirow{2}{*}{Methods} & \multirow{2}{*}{Modality} & \multicolumn{5}{c|}{ADNI1 \cite{ADNI}} & \multicolumn{5}{c|}{ADNI2 \cite{ADNI}} \\ \cline{3-12}
& & ACC & SPE & SEN & AUC & F1 & ACC & SPE & SEN & AUC & F1 \\ \hline
ResNet \cite{ResNet} & \textbf{M+P} & 0.725 & 0.823& 0.564  &  0.653  & 0.606   & 0.809 & 0.928 & 0.437 & 0.709 & 0.510 \\ \hline
JSRL \cite{JSRL}    & \textbf{M+P} & 0.582 & 0.779 & 0.354 & 0.571 & 0.580 & 0.650 & 0.590 & 0.720 & 0.694 & 0.602 \\ \hline
HOPE \cite{hope}  & \textbf{M} & 0.611  &  0.786 & 0.611 &  0.648 & 0.593 & 0.712 & 0.860 & 0.712 & 0.616 & 0.692 \\ \hline
VAPL \cite{VAPL}  & \textbf{M+C} & 0.630 & 0.564 & 0.693 & 0.635 & 0.651 & 0.835 & 0.843 &0.745  & 0.865 & 0.671 \\ \hline
HFBSurv \cite{li2022hfbsurv}  & \textbf{M+P+C} &0.921   &0.904  &0.937   &0.920   &0.916   &0.954   &0.977   &0.909   &0.943   &0.932   \\ \hline
ITCFN (w/o MMG) & \textbf{M+P+C} &0.932   &0.925   &0.937   &0.931  &0.927   &\textbf{0.960}   &\textbf{0.992}   &\textbf{0.937}   &\textbf{0.965}   &\textbf{0.960}  \\ \hline
\textbf{ ITCFN (Ours)} & \textbf{M+P+C} & \textbf{0.947} & \textbf{0.949} & \textbf{0.944} & \textbf{0.946} & \textbf{0.944} & 0.954 & 0.980 & 0.906 &0.943 & 0.931 \\ \hline
\end{tabular}
\label{comparison}
\vspace{-1em}

\end{table*}

\begin{table*}[t]

\caption{Ablation performance with and without enhancement module.}
\centering
\small
\begin{tabular}{|p{2.3cm}<{\centering}|p{1cm}<{\centering}|p{1cm}<{\centering}|p{1cm}<{\centering}|p{1cm}<{\centering}|p{1cm}<{\centering}|p{1cm}<{\centering}|p{1cm}<{\centering}|p{1cm}<{\centering}|p{1cm}<{\centering}|p{1cm}<{\centering}|p{1cm}<{\centering}|}
\hline
\multirow{2}{*}{Methods} & \multicolumn{5}{c|}{ADNI1 \cite{ADNI}} & \multicolumn{5}{c|}{ADNI2 \cite{ADNI}} \\ \cline{2-11}
& ACC & SPE & SEN & AUC & F1 & ACC & SPE & SEN & AUC & F1 \\ \hline
None    &0.889   &0.904   &0.877   &0.904   &0.895   &0.941   &0.963   &0.918   &0.932   &0.934   \\ \hline
MMG only &0.889   &0.890  &0.891  &0.890   &0.879   &0.948   &0.976   &0.914   &0.935   &0.932   \\ \hline
TCAF only &0.932   &0.925   &0.937   &0.931  &0.927 &\textbf{0.960}   &\textbf{0.992}   &\textbf{0.937}   &\textbf{0.965}   &\textbf{0.960}   \\ \hline
MMG+TCAF  & \textbf{0.947} & \textbf{0.949} & \textbf{0.944} & \textbf{0.946} & \textbf{0.944} & 0.954 & 0.980 & 0.906 & 0.943 & 0.931 \\ \hline
\end{tabular}
\label{Ablation}
\vspace{-0.5em}

\end{table*}

\section{EXPERIMENTAL RESULTS}
\label{sec:typestyle}
\subsection{Experimental Setup}
\textbf{Dataset.} The dataset for this study is obtained from the Alzheimer's Disease Neuroimaging Initiative (ADNI), specifically the ADNI-1 and ADNI-2 cohorts \cite{ADNI}. To prevent duplication, subjects present in both datasets were removed from ADNI-2. We selected T1-weighted sMRI, FDG-PET, and clinical data, categorized into four groups: normal controls (NC), sMCI, pMCI, and AD. Demographic information of the dataset is shown in Table \ref{dataset}. 


All MRI images underwent preprocessing steps, including intensity normalization, skull stripping, and normalization to MNI space. FDG-PET images were processed with intensity normalization, MNI space normalization, and co-registration with MRI. For clinical data, we selected seven attributes: age, gender, education, ApoE4 status, P-tau 181, T-tau, and summary measures from 18F-FDG PET imaging.


\textbf{Metrics.} 
We employed multiple metrics to assess the effectiveness and robustness of the ITCFN model, including accuracy (ACC), sensitivity (SEN), specificity (SPE), the Area Under the Curve (AUC), and the F1 score (F1).

\subsection{Implementation Details}
\vspace{-0.5em}
To ensure reproducible and comparable results, we employed 5-fold cross-validation in all experiments, validating the model's stability and generalization while maintaining a consistent random seed for data splitting. We conducted our experiments using the PyTorch 2.0 framework, utilizing a single NVIDIA A100 80 GB GPU for computational efficiency. The model was trained from scratch over two distinct stages, each consisting of 200 epochs, with a batch size of 8 to effectively manage the data. We optimized the model parameters using the Adam algorithm, setting the learning rate to 0.0001 to ensure precise adjustments during training.

\vspace{-1em}
\subsection{Comparative Experiments}
\vspace{-0.5em}
Table \ref{comparison} shows the comparative results of different methods against our proposed ITCFN. In the Modality column, \textbf{M} stands for MRI, \textbf{P} for PET, and \textbf{C} for clinical data. The HOPE method achieves some recognition performance by utilizing features from a single modality (MRI) linked to AD progression. Although ResNet performs effectively, it struggles to combine multiple modalities, which limits its performance with limited and imbalanced data. The JSRL method addresses the issue of missing modalities but performs poorly on small datasets, likely because it directly concatenates the modalities. In contrast, the VPAL method enhances feature representation by integrating heterogeneous data. Finally, HFBSurv effectively merges tri-modal features using a factorized bilinear model, allowing for the progressive fusion of multimodal data and thus achieving superior performance.

Our method surpasses all others across all metrics. On the ADNI1 dataset, our framework reaches an ACC of 0.947, SPE of 0.949, SEN of 0.944, AUC of 0.946, and F1 of 0.944, improving by 0.026, 0.045, 0.007, 0.026, and 0.028 over the best results from other methods, respectively. Similarly, on the ADNI2 dataset, our approach achieves an ACC of 0.960, SPE of 0.992, SEN of 0.937, AUC of 0.965, and F1 of 0.960, with enhancements of 0.006, 0.015, 0.028, 0.022, and 0.028 compared to the best results of other methods.

\vspace{-1em}
\subsection{Ablation Study}
\vspace{-0.5em}
We assessed the effectiveness of the MMG and TCAF modules in addressing issues of missing data and modality fusion. In the ADNI1 dataset, where missing data is a significant challenge, the MMG module enhanced the model's ability to recognize pMCI, increasing SEN by 0.014. The TCAF module complemented this by effectively integrating information from different modalities, improving ACC by 0.043 and the AUC by 0.027. In the ADNI2 dataset, characterized by a greater data imbalance, the MMG module generated 30 sMCI cases but only 1 pMCI case, somewhat worsening the imbalance. Despite this, the TCAF module achieved optimal performance on its own, enhancing ACC by 0.019 and AUC by 0.033. Nevertheless, the MMG module still contributed to improving the model's overall performance.




\vspace{-1em}

\section{Conclusion}
\vspace{-0.5em}
In this study, we present an innovative method known as ITCFN to predict the conversion of MCI. The ITCFN method combines MRI, PET, and clinical features, effectively managing missing PET data through the MMG module and achieving modality fusion via the TCAF module. This model exhibits superior classification performance when compared to traditional unimodal methods and other multimodal deep learning techniques. Through ablation studies and comparative analyses, we confirm the effectiveness of both the MMG and TCAF modules. However, our approach has certain limitations. Currently, we only consider baseline MRI. Incorporating longitudinal MRI data could provide deeper insights into the progression of the disease. Additionally, our model does not include genetic data, which might further enhance its predictive accuracy.
\section{Compliance with Ethical Standards}
This study retrospectively utilized human subject data obtained from the publicly accessible ADNI dataset \cite{ADNI}. The license accompanying this open-access data confirms that no ethical approval is required.
\section{Acknowledgements}
This work was supported by the Open Project Program of the State Key Laboratory of CAD\&CG, Zhejiang University (Grant No.A2410), National Natural Science Foundation of China (No.61702146, 62076084), Guangxi Key R\&D Project (No.AB24010167), GuangDong Basic and Applied Basic Research Foundation (No.2022A1515110570), Shenzhen Longgang District Science and Technology Innovation Special Fund (No. LGKCYLWS2023018).
\bibliographystyle{IEEEbib}
\bibliography{refs_simple}
\end{document}